\documentclass[%
 aip,
 apl,
 amsmath,amssymb,
 reprint,%
]{revtex4-1}

\usepackage{graphicx}
\usepackage{dcolumn}
\usepackage{bm}
\usepackage[version=4]{mhchem}
\usepackage[utf8]{inputenc}
\usepackage[T1]{fontenc}
\usepackage{mathptmx}
\usepackage{etoolbox}
\usepackage{natbib}
\usepackage{siunitx}
\newcommand{\gtwo}{$g^{(2)}\left(\tau\right)$ }

\makeatletter
\def\@email#1#2{%
 \endgroup
 \patchcmd{\titleblock@produce}
  {\frontmatter@RRAPformat}
  {\frontmatter@RRAPformat{\produce@RRAP{*#1\href{mailto:#2}{#2}}}\frontmatter@RRAPformat}
  {}{}
}%
\makeatother
\begin{document}

\preprint{AIP/123-QED}

\title[Direct-write projection lithography of quantum dot micropillar single photon sources]{Direct-write projection lithography of quantum dot micropillar single photon sources}

\author{Petros Androvitsaneas}
\affiliation{School of Engineering, Cardiff University, Queen's Building, The Parade, Cardiff, CF24 3AA, UK.}
\affiliation{Translational Research Hub, Maindy Road, Cardiff, CF24 4HQ, UK}

\author{Rachel N. Clark}
\affiliation{School of Engineering, Cardiff University, Queen's Building, The Parade, Cardiff, CF24 3AA, UK.}
\affiliation{Translational Research Hub, Maindy Road, Cardiff, CF24 4HQ, UK}

\author{Matthew Jordan}
\affiliation{School of Engineering, Cardiff University, Queen's Building, The Parade, Cardiff, CF24 3AA, UK.}
\affiliation{Translational Research Hub, Maindy Road, Cardiff, CF24 4HQ, UK}

\author{Tomas Peach}
\affiliation{Institute for Compound Semiconductors, Cardiff University, Queen's Buildings, The Parade, Cardiff, CF24 3AA, UK.}

\author{Stuart Thomas}
\affiliation{Institute for Compound Semiconductors, Cardiff University, Queen's Buildings, The Parade, Cardiff, CF24 3AA, UK.}

\author{Saleem Shabbir}
\affiliation{Institute for Compound Semiconductors, Cardiff University, Queen's Buildings, The Parade, Cardiff, CF24 3AA, UK.}

\author{Angela D. Sobiesierski}
\affiliation{Institute for Compound Semiconductors, Cardiff University, Queen's Buildings, The Parade, Cardiff, CF24 3AA, UK.}

\author{Aristotelis Trapalis}
\affiliation{Department of Electronic and Electrical Engineering, University of Sheffield, Mappin Street, S1 3JD, Sheffield, UK.}

\author{Ian A. Farrer}
\affiliation{Department of Electronic and Electrical Engineering, University of Sheffield, Mappin Street, S1 3JD, Sheffield, UK.}
\affiliation{EPSRC National Epitaxy Facility, University of Sheffield, Sheffield, S3 7HQ, UK.}

\author{Wolfgang W. Langbein}
\affiliation{School of Physics and Astronomy, Cardiff University, Queen's Building, The Parade, Cardiff, UK, CF24 3AA, UK.}

\author{Anthony J. Bennett}
\email{BennettA19@cardiff.ac.uk}
\affiliation{School of Engineering, Cardiff University, Queen's Building, The Parade, Cardiff, CF24 3AA, UK.}
\affiliation{Translational Research Hub, Maindy Road, Cardiff, CF24 4HQ, UK}
\affiliation{School of Physics and Astronomy, Cardiff University, Queen's Building, The Parade, Cardiff, UK, CF24 3AA, UK.}

\date{\today}

\begin{abstract}
We have developed a process to mass-produce quantum dot micropillar cavities using direct-write lithography. This technique allows us to achieve high volume patterning of  high aspect ratio pillars with vertical, smooth sidewalls maintaining a high quality factor for diameters below \SI{2.0}{\micro \metre}. Encapsulating the cavities in a thin layer of oxide (Ta$_2$O$_5$) prevents oxidation in the atmosphere, preserving the optical properties of the cavity over months of ambient exposure. We confirm that single dots in the cavities can be deterministically excited to create high purity indistinguishable single photons with interference visibility $(96.2\pm0.7)\%$.
\end{abstract}

\maketitle

Single photon sources are an essential building block for a variety of quantum technologies~\cite{arakawa_rev}. Developments in resonant excitation~\cite{Flagg2009,SomaschiN.:2016uq}, in-situ lithography~\cite{PhysRevLett.101.267404,SomaschiN.:2016uq} and cavity design~\cite{Tomm2021,He2013} have made quantum dots (QDs) one of the main contenders for high efficiency and high indistinguishablity single photon sources. Furthermore, the potential to entangle photons sequentially emitted by the QDs using spin opens up new functionality in entangled photon pair generation~\cite{Stevenson2006,Young_2006,Dousse2010,PhysRevLett.123.160501}, cluster state generation~\cite{doi:10.1126/science.aah4758} and other higher-dimensional photonic states~\cite{Lee_2019,PhysRevLett.128.233602}. 

One of the most promising cavity designs is the semiconductor micropillar cavity ~\cite{SomaschiN.:2016uq,Ding:2016fk,Wang2019} in which two distributed Bragg reflectors (DBR) surround a spacer layer containing a low density layer of quantum dots. When etched into circular pillars of approximately \SI{2}{\micro\metre} these structures confine localised optical modes that enhance photon emission from the QDs, whilst coupling efficiently to a Gaussian mode that can be collected in the far field. A key challenge is achieving a deep vertical etch; this requires balancing the chemical and mechanical properties of the etch to manage the rate of re-deposition and minimise damage to the mask layer. Furthermore, the etched pillar sides must be smooth to limit the scattering loss and maintain a high quality factor ($Q$) and light collection efficiency. Therefore, fabrication requires a hard mask able to withstand the aggressive etch required to remove up to \SI{10}{\micro\metre} of semiconductor, but which is thin enough to be patterned with high accuracy is required. Different approaches to masking for this purpose have been demonstrated, including randomly positioned sapphire nanocrystals~\cite{Santori2002}, contact lithography with a quartz mask~\cite{Bennett2016},  electron beam lithography~\cite{Schneider2016} and cryogenic in-situ laser-lithography~\cite{PhysRevLett.101.267404}. The latter two allow for pre-selection of promising QDs and alignment of cavities, but are expensive and less compatible with mass production of devices.    

Here we report a direct-write photolithography method allowing high-throughput sample patterning for deep etches of GaAs. This technique, also known as mask-less lithography, uses a UV light source and a digital light modulator to project the pattern onto the sample, with potential to reconfigure designs by software. It provides the flexibility of electron beam lithography, at a lower cost, and with a \SI{400}{\nano\metre} resolution is sufficient for this application. After etching, we encapsulate the sample in a few nanometer thick oxide layer (C$_4$F$_8$/O$_2$), protecting against oxidation in the atmosphere. Characterisation of the cavities show they have low sidewall scattering parameters, retaining high $Q$ even at low diameter. Finally, we demonstrate a high brightness, high purity and indistinguishable single photon source using deterministic pulsed resonant excitation, to verify the quality of the material.

\begin{figure}
\includegraphics[width=8.5cm]{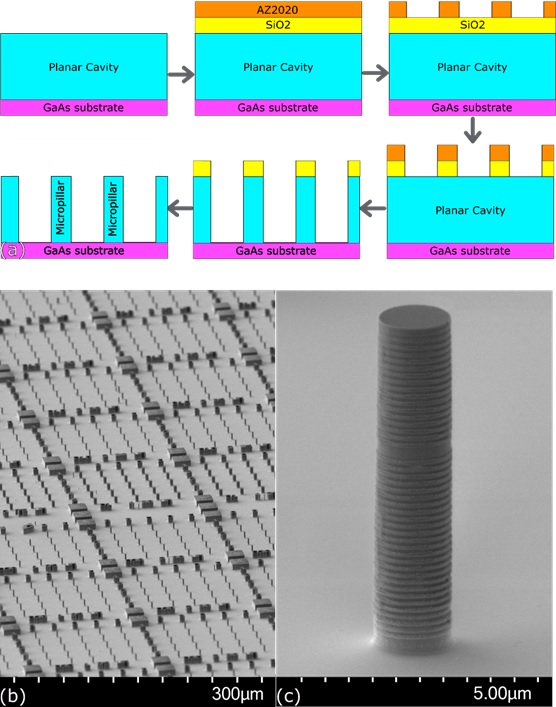}
\caption{\label{fig:micro} Micropillar fabrication. (a) Schematic of processing steps. (b) Wide area scanning electron microscope image (SEM) of etched structures with a variety of different diameters. (c) SEM of a high $Q$-factor micropillar of diameter \SI{1.75} {\micro \metre}}
\end{figure}

The samples were grown by molecular beam epitaxy. A high-$Q$ cavity sample was grown consisting of a lower Bragg mirror of 26 pairs of alternating Al$_{0.95}$Ga$_{0.05}$As~\cite{CHO20101955} and GaAs  $\lambda/4$ layers, a single wavelength spacer with InAs QDs at its center, and a final 17 pair Bragg mirror. A low-$Q$ cavity was also grown with 7/26 Bragg mirror pairs. The design wavelength  was \SI{940} {\nano \metre}. The processing proceeds as shown in Fig. \ref{fig:micro} (a) by coating the chips with a hard-mask layer of \SI{750} {\nano \metre} \ce{SiO2} deposited via plasma enhanced chemical vapour deposition (PECVD). A \SI{2}{\micro\metre} layer of negative photo-resist, AZ2020, is applied and exposed using the MicroWriter ML3 Pro direct-write photo-lithography tool. The pattern consists of discs with diameters in the range \SI{1.55} {\micro \metre} to \SI{5.00} {\micro \metre} in regularly spaced $5\times5$ arrays. This direct-write method allows for the patterning of $14,000$ devices in  \SI{240} {\second}. After developing the photoresist in AZ726, the hard mask is etched using a C$_4$F$_8$/O$_2$ inductively coupled plasma (ICP) and the photo-resist removed. The semiconductor is then etched using a Cl$_2$/BCl$_3$/N$_2$ ICP. The hard mask is then removed with a second C$_4$F$_8$/O$_2$ etch. Finally, the micropillars are encapsulated in a \SI{10} {\nano \metre} layer of Ta$_2$O$_5$ using atomic layer deposition. This oxide layer provides a uniform conformal coating that protects the samples against oxidation, especially for the DBR layers containing aluminium~\cite{Tomm2021}.

The quality of the patterning and semiconductor etch determines the sidewall roughness, which introduces losses to the cavity mode HE$_{11}$~\cite{Schneider2016}. The overall loss rate for photons in the mode is inversely proportional to the quality factor at a given diameter, $1/Q(d) =1/Q_0+1/Q_s(d)$, where the decay rate due to sidewall roughness is parameterized by $1/Q_s(d)$, which adds to the loss rate through mirrors $1/Q_0$. $Q_0$ can be determined from the $Q$-factor that cavities tend towards at high diameter. $Q_s(d)$ is linked to the diameter of the micropillar by the following expression $1/Q_s(d)=2 k_sJ_0^2(kd/2)/d$, where $k_s$ is the sidewall loss coefficient, $J_0(kd/2)$ is the $0$\textsuperscript{th} order Bessel function with $k$ the transverse wavevector and $d$ the diameter~\cite{Schneider2016}.

\begin{figure}
\includegraphics[width=8.5cm]{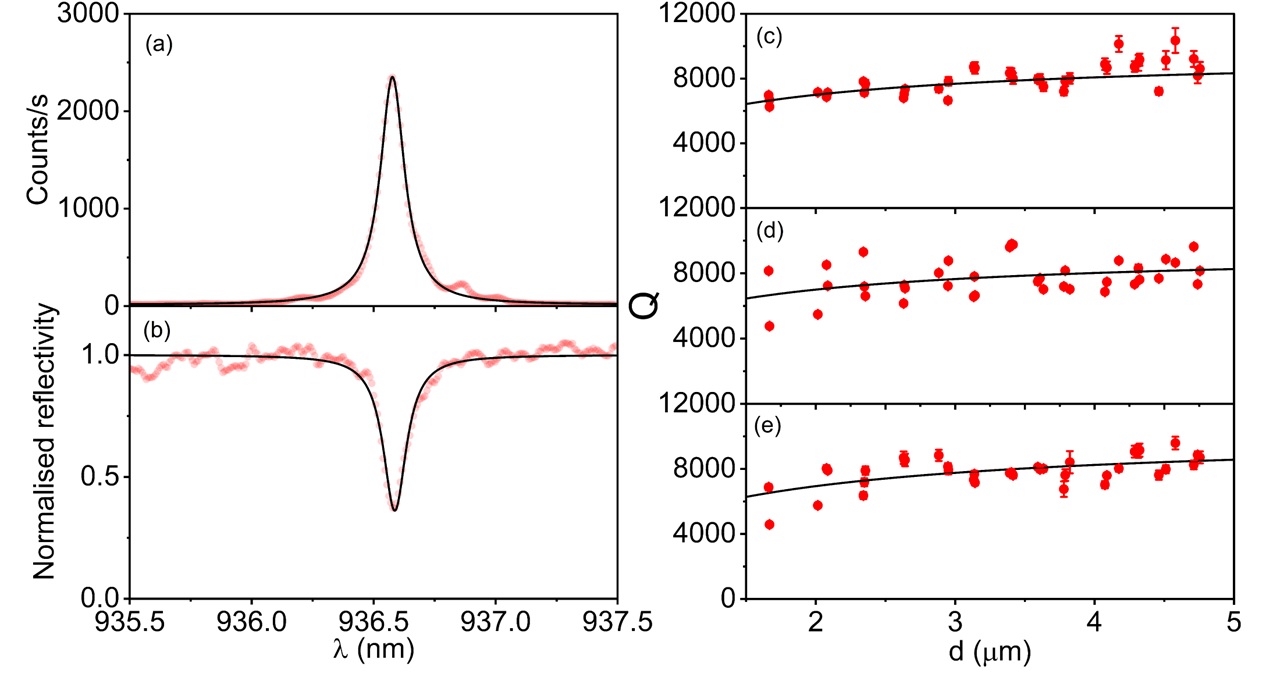}
\caption{\label{fig:Qfactor} Cavity quality factors. (a) Photoluminescence (PL) spectrum of the cavity mode (red data points) with the corresponding Lorentzian fit (black line) of a micropillar consisting of 17/26 period Bragg mirrors and a diameter $d=$\SI{3.14} {\micro \metre} at \SI{80} {\kelvin}, yielding a $Q=7740\pm60$. (b) Normalised white light reflectivity (WLR) (red data points) for the same micropillar at  \SI{80} {\kelvin} with the Lorentzian fit (black  line), yielding a $Q=8100\pm300$. (c) $Q(d)$ measured at \SI{4} {\kelvin} by WLR (d) $Q(d)$ measured at \SI{80} {\kelvin} by PL and (e) by WLR. Fits shown as black lines discussed in the text. }
\end{figure}

Two different techniques have been utilised to measure the cavity's $Q(d)$, photoluminescence (PL) and white light reflectivity (WLR), with example data shown in Fig.~\ref{fig:Qfactor}(a) and (b). The relatively high density of spectrally sharp QD transitions in the spectral range of the mode, made the measurement of $Q$ using PL at \SI{4} {\kelvin} unreliable. Therefore, the WLR measurement was used to determine the $Q$-factor of HE$_{11}$ at this temperature,  Fig.~\ref{fig:Qfactor}(c). Additionally, we measure the $Q$-factors at \SI{80} {\kelvin} using PL (Fig.~\ref{fig:Qfactor}(d)) and WLR reflectivity (Fig. \ref{fig:Qfactor}(e)). All three datasets yield a similar value for the sidewall loss coefficient $k_s$, (c) $k_s$ of $(48\pm9)$~\si{\pico\metre}, (d) $(50\pm20)$~\si{\pico\metre}  and (e) $(60\pm20)$~\si{\pico\metre}. These values are comparable to the state-of-the-art for these photonic structures~\cite{Schneider2016} which reports $k_s = 68$~\si{\pico\metre}. 

\begin{figure*}
\includegraphics[width=17.5cm]{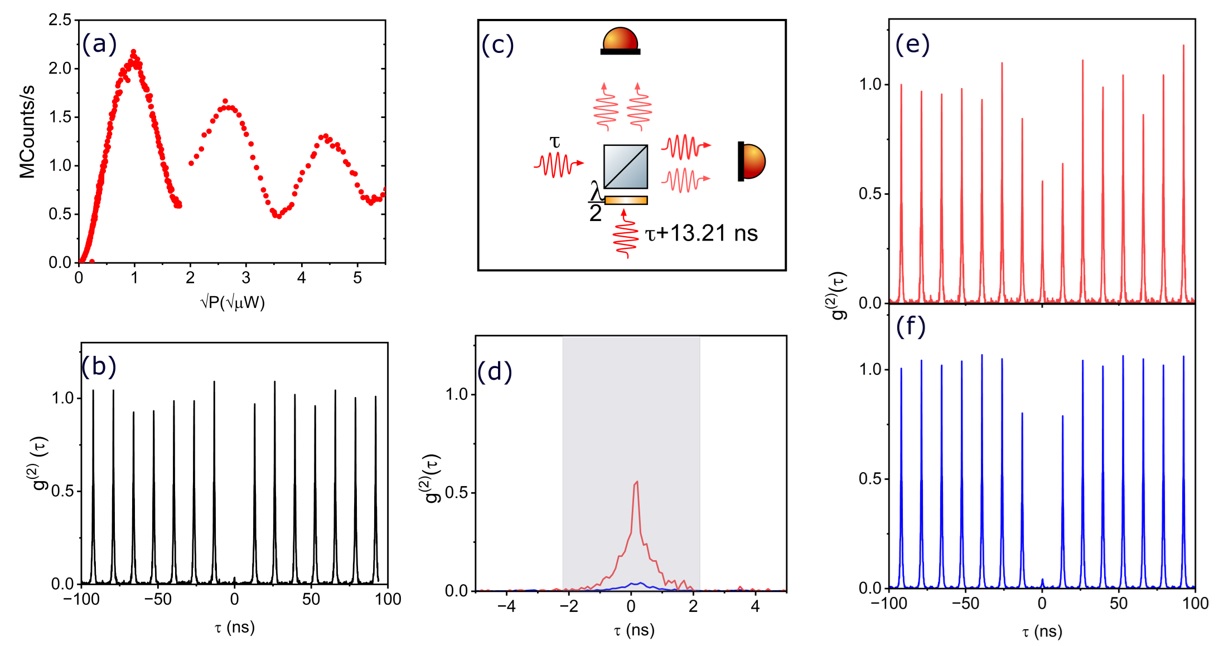}
\caption{\label{fig:hom}Indistinguishable single photons from a deterministically driven neutral exciton transition in a micropillar. (a) Rabi oscillation in pulse amplitude for a neutral exciton in a 7/26 period Bragg mirror cavity, the discontinuity at the point of $2\pi$ stems from the measurement using a neutral density filter to allow the two power ranges. (b) Pulsed \gtwo produced by fully inverting the system using a $\pi$-pulse, at a power of \SI{0.961}{\nano \watt}, with $g^{(2)}(0)=0.027\pm0.004$. (c) Setup used to interfere two subsequent photons generated \SI{13.2}{\nano \second} apart. A half waveplate ($\lambda/2$) is used to introduce polarization indistinguishability. (d) Hong-Ou-Mandel interference measurement for orthogonally polarized photons (red) and co-polarized photons (blue). The visibility is $(88.9\pm0.6)\%$, derived by the ratio of the areas denoted by the shaded rectangle. Accounting for imperfections in the interferometer and the probability of multi-photon emission from the source, the inferred two photon interference is $(96.2\pm0.7)\%$. (e) Correlation for orthogonally polarized photons. (f) Correlation for co-polarized photons.}
\end{figure*}

We then test the quantum light emission from these samples under resonant excitation. The sample was stored in air for three months after the processing with no observable degradation in $Q$. We study a neutral exciton on resonance with the HE$_{11}$ in a \SI{1.7}{\micro \metre} diameter micropillar with a low $Q$-factor ($Q=440$) and 7/26 Bragg mirror pairs. We focus on this sample because every pillar has transitions within broad cavity mode which can be resonantly excited. It has been show that these low $Q$ cavities can be efficient and broadband ~\cite{Androvitsaneas:2016kx,PhysRevLett.129.033601}. The radiative lifetime of the transition is $T_1=$~\SI{480}{\pico \second} with a fine structure splitting of \SI{11}{\micro\eV}. Simulations suggest the maximum expected Purcell factor we could observe would be $3.33$, with an overall expected efficiency of $0.73$ at the collection objective. Exciting the transition resonantly in a cross-polarized geometry~\cite{kuhlmandark} we vary the pulse amplitude to observe a Rabi oscillation. With $\pi$-pulse excitation the maximum count-rate is $\sim2.2$~MHz, see Fig.~\ref{fig:hom}(a). A second order autocorrelation measurement reveals the purity of the single photon emission~\cite{hbt}, when the system undergoes a full population inversion under the excitation of $\pi$-pulse, to give $g^{(2)} (0)=0.027\pm0.004$, see Fig.~\ref{fig:hom}(b). 

Finally, we measure the indistinguishability of the single photons emitted under the conditions described above by interfering two sequentially emitted photons  from the QD, Fig.~\ref{fig:hom}(c). This yields a raw visibility of the two-photon interference $(88.9\pm0.6)\%$ Fig.~\ref{fig:hom}(d-e). Based on the interferometer visibility and the value of the $g^{(2)}(0)$ the overall corrected two photon interference visibility corresponds to $(96.2\pm0.7)\%$~\cite{hom_cor}. This value is comparable to the visibility achieved with QDs in cavities with higher Purcell factors~\cite{arakawa_rev,SomaschiN.:2016uq,Ding:2016fk,Wang2019,Tomm2021}. This shows the excellent condition of the material even after the prolonged exposure to a non-controlled atmosphere.

This direct-write method can be used for high volume manufacturing of QD micropillar devices. The quality of the structure, low sidewall roughness, and high purity of the indistinguishable photons, shows its promise as a flexible platform for mass production of  single photon sources. Future work could improve the collection efficiency into a single mode fiber by optimizing the far field emission pattern. An increased yield of optimized structures could be achieved by mapping the locations of dots prior to the processing, facilitating the repositioning of cavities and dots over a whole chip without the need for cryogenic lithography ~\cite{PhysRevLett.101.267404}. Furthermore, with positioned arrays of QDs becoming available~\cite{doi:10.1063/5.0013718}, the yield could approach unity. 

\begin{acknowledgments}
We acknowledge financial support provided by EPSRC via Grant No. EP/T017813/1 and EP/T001062/1. RC was supported by grant EP/S024441/1, Cardiff University and the National Physical Laboratory(NPL). We further thank Dr Alastair Sinclair and Dr Philip Dolan at NPL for technical discussions. Device processing was carried out in the cleanroom of the ERDF-funded Institute for Compound Semiconductors (ICS) at Cardiff University.
For the purpose of open access, the author has applied a CC BY public copyright licence.
\end{acknowledgments}

\section*{Data Availability Statement}

The data that support the findings of this study are available upon request.

\section*{References}

\bibliography{bib13deskpetros}

\begin{thebibliography}{25}%
\makeatletter
\providecommand \@ifxundefined [1]{%
 \@ifx{#1\undefined}
}%
\providecommand \@ifnum [1]{%
 \ifnum #1\expandafter \@firstoftwo
 \else \expandafter \@secondoftwo
 \fi
}%
\providecommand \@ifx [1]{%
 \ifx #1\expandafter \@firstoftwo
 \else \expandafter \@secondoftwo
 \fi
}%
\providecommand \natexlab [1]{#1}%
\providecommand \enquote  [1]{``#1''}%
\providecommand \bibnamefont  [1]{#1}%
\providecommand \bibfnamefont [1]{#1}%
\providecommand \citenamefont [1]{#1}%
\providecommand \href@noop [0]{\@secondoftwo}%
\providecommand \href [0]{\begingroup \@sanitize@url \@href}%
\providecommand \@href[1]{\@@startlink{#1}\@@href}%
\providecommand \@@href[1]{\endgroup#1\@@endlink}%
\providecommand \@sanitize@url [0]{\catcode `\\12\catcode `\$12\catcode
  `\&12\catcode `\#12\catcode `\^12\catcode `\_12\catcode `\%12\relax}%
\providecommand \@@startlink[1]{}%
\providecommand \@@endlink[0]{}%
\providecommand \url  [0]{\begingroup\@sanitize@url \@url }%
\providecommand \@url [1]{\endgroup\@href {#1}{\urlprefix }}%
\providecommand \urlprefix  [0]{URL }%
\providecommand \Eprint [0]{\href }%
\providecommand \doibase [0]{http://dx.doi.org/}%
\providecommand \selectlanguage [0]{\@gobble}%
\providecommand \bibinfo  [0]{\@secondoftwo}%
\providecommand \bibfield  [0]{\@secondoftwo}%
\providecommand \translation [1]{[#1]}%
\providecommand \BibitemOpen [0]{}%
\providecommand \bibitemStop [0]{}%
\providecommand \bibitemNoStop [0]{.\EOS\space}%
\providecommand \EOS [0]{\spacefactor3000\relax}%
\providecommand \BibitemShut  [1]{\csname bibitem#1\endcsname}%
\let\auto@bib@innerbib\@empty
\bibitem [{\citenamefont {Arakawa}\ and\ \citenamefont
  {Holmes}(2020)}]{arakawa_rev}%
  \BibitemOpen
  \bibfield  {author} {\bibinfo {author} {\bibfnamefont {Y.}~\bibnamefont
  {Arakawa}}\ and\ \bibinfo {author} {\bibfnamefont {M.~J.}\ \bibnamefont
  {Holmes}},\ }\href@noop {} {\bibfield  {journal} {\bibinfo  {journal}
  {Applied Physics Reviews}\ }\textbf {\bibinfo {volume} {7}},\ \bibinfo
  {pages} {021309} (\bibinfo {year} {2020})}\BibitemShut {NoStop}%
\bibitem [{\citenamefont {Flagg}\ \emph {et~al.}(2009)\citenamefont {Flagg},
  \citenamefont {Muller}, \citenamefont {Robertson}, \citenamefont {Founta},
  \citenamefont {Deppe}, \citenamefont {Xiao}, \citenamefont {Ma},
  \citenamefont {Salamo},\ and\ \citenamefont {Shih}}]{Flagg2009}%
  \BibitemOpen
  \bibfield  {author} {\bibinfo {author} {\bibfnamefont {E.~B.}\ \bibnamefont
  {Flagg}}, \bibinfo {author} {\bibfnamefont {A.}~\bibnamefont {Muller}},
  \bibinfo {author} {\bibfnamefont {J.}~\bibnamefont {Robertson}}, \bibinfo
  {author} {\bibfnamefont {S.}~\bibnamefont {Founta}}, \bibinfo {author}
  {\bibfnamefont {D.}~\bibnamefont {Deppe}}, \bibinfo {author} {\bibfnamefont
  {M.}~\bibnamefont {Xiao}}, \bibinfo {author} {\bibfnamefont {W.}~\bibnamefont
  {Ma}}, \bibinfo {author} {\bibfnamefont {G.}~\bibnamefont {Salamo}}, \ and\
  \bibinfo {author} {\bibfnamefont {C.-K.}\ \bibnamefont {Shih}},\ }\href@noop
  {} {\bibfield  {journal} {\bibinfo  {journal} {Nature Physics}\ }\textbf
  {\bibinfo {volume} {5}},\ \bibinfo {pages} {203} (\bibinfo {year}
  {2009})}\BibitemShut {NoStop}%
\bibitem [{\citenamefont {Somaschi}\ \emph {et~al.}(2016)\citenamefont
  {Somaschi}, \citenamefont {Giesz}, \citenamefont {De~Santis}, \citenamefont
  {Loredo}, \citenamefont {Almeida}, \citenamefont {Hornecker}, \citenamefont
  {Portalupi}, \citenamefont {Grange}, \citenamefont {Anton}, \citenamefont
  {Demory} \emph {et~al.}}]{SomaschiN.:2016uq}%
  \BibitemOpen
  \bibfield  {author} {\bibinfo {author} {\bibfnamefont {N.}~\bibnamefont
  {Somaschi}}, \bibinfo {author} {\bibfnamefont {V.}~\bibnamefont {Giesz}},
  \bibinfo {author} {\bibfnamefont {L.}~\bibnamefont {De~Santis}}, \bibinfo
  {author} {\bibfnamefont {J.}~\bibnamefont {Loredo}}, \bibinfo {author}
  {\bibfnamefont {M.~P.}\ \bibnamefont {Almeida}}, \bibinfo {author}
  {\bibfnamefont {G.}~\bibnamefont {Hornecker}}, \bibinfo {author}
  {\bibfnamefont {S.~L.}\ \bibnamefont {Portalupi}}, \bibinfo {author}
  {\bibfnamefont {T.}~\bibnamefont {Grange}}, \bibinfo {author} {\bibfnamefont
  {C.}~\bibnamefont {Anton}}, \bibinfo {author} {\bibfnamefont
  {J.}~\bibnamefont {Demory}},  \emph {et~al.},\ }\href@noop {} {\bibfield
  {journal} {\bibinfo  {journal} {Nature Photonics}\ }\textbf {\bibinfo
  {volume} {10}},\ \bibinfo {pages} {340} (\bibinfo {year} {2016})}\BibitemShut
  {NoStop}%
\bibitem [{\citenamefont {Dousse}\ \emph {et~al.}(2008)\citenamefont {Dousse},
  \citenamefont {Lanco}, \citenamefont {Suffczy{\'n}ski}, \citenamefont
  {Semenova}, \citenamefont {Miard}, \citenamefont {Lema{\^\i}tre},
  \citenamefont {Sagnes}, \citenamefont {Roblin}, \citenamefont {Bloch},\ and\
  \citenamefont {Senellart}}]{PhysRevLett.101.267404}%
  \BibitemOpen
  \bibfield  {author} {\bibinfo {author} {\bibfnamefont {A.}~\bibnamefont
  {Dousse}}, \bibinfo {author} {\bibfnamefont {L.}~\bibnamefont {Lanco}},
  \bibinfo {author} {\bibfnamefont {J.}~\bibnamefont {Suffczy{\'n}ski}},
  \bibinfo {author} {\bibfnamefont {E.}~\bibnamefont {Semenova}}, \bibinfo
  {author} {\bibfnamefont {A.}~\bibnamefont {Miard}}, \bibinfo {author}
  {\bibfnamefont {A.}~\bibnamefont {Lema{\^\i}tre}}, \bibinfo {author}
  {\bibfnamefont {I.}~\bibnamefont {Sagnes}}, \bibinfo {author} {\bibfnamefont
  {C.}~\bibnamefont {Roblin}}, \bibinfo {author} {\bibfnamefont
  {J.}~\bibnamefont {Bloch}}, \ and\ \bibinfo {author} {\bibfnamefont
  {P.}~\bibnamefont {Senellart}},\ }\href@noop {} {\bibfield  {journal}
  {\bibinfo  {journal} {Physical review letters}\ }\textbf {\bibinfo {volume}
  {101}},\ \bibinfo {pages} {267404} (\bibinfo {year} {2008})}\BibitemShut
  {NoStop}%
\bibitem [{\citenamefont {Tomm}\ \emph {et~al.}(2021)\citenamefont {Tomm},
  \citenamefont {Javadi}, \citenamefont {Antoniadis}, \citenamefont {Najer},
  \citenamefont {L{\"o}bl}, \citenamefont {Korsch}, \citenamefont {Schott},
  \citenamefont {Valentin}, \citenamefont {Wieck}, \citenamefont {Ludwig} \emph
  {et~al.}}]{Tomm2021}%
  \BibitemOpen
  \bibfield  {author} {\bibinfo {author} {\bibfnamefont {N.}~\bibnamefont
  {Tomm}}, \bibinfo {author} {\bibfnamefont {A.}~\bibnamefont {Javadi}},
  \bibinfo {author} {\bibfnamefont {N.~O.}\ \bibnamefont {Antoniadis}},
  \bibinfo {author} {\bibfnamefont {D.}~\bibnamefont {Najer}}, \bibinfo
  {author} {\bibfnamefont {M.~C.}\ \bibnamefont {L{\"o}bl}}, \bibinfo {author}
  {\bibfnamefont {A.~R.}\ \bibnamefont {Korsch}}, \bibinfo {author}
  {\bibfnamefont {R.}~\bibnamefont {Schott}}, \bibinfo {author} {\bibfnamefont
  {S.~R.}\ \bibnamefont {Valentin}}, \bibinfo {author} {\bibfnamefont {A.~D.}\
  \bibnamefont {Wieck}}, \bibinfo {author} {\bibfnamefont {A.}~\bibnamefont
  {Ludwig}},  \emph {et~al.},\ }\href@noop {} {\bibfield  {journal} {\bibinfo
  {journal} {Nature Nanotechnology}\ }\textbf {\bibinfo {volume} {16}},\
  \bibinfo {pages} {399} (\bibinfo {year} {2021})}\BibitemShut {NoStop}%
\bibitem [{\citenamefont {He}\ \emph {et~al.}(2013)\citenamefont {He},
  \citenamefont {He}, \citenamefont {Wei}, \citenamefont {Wu}, \citenamefont
  {Atat{\"u}re}, \citenamefont {Schneider}, \citenamefont {H{\"o}fling},
  \citenamefont {Kamp}, \citenamefont {Lu},\ and\ \citenamefont
  {Pan}}]{He2013}%
  \BibitemOpen
  \bibfield  {author} {\bibinfo {author} {\bibfnamefont {Y.-M.}\ \bibnamefont
  {He}}, \bibinfo {author} {\bibfnamefont {Y.}~\bibnamefont {He}}, \bibinfo
  {author} {\bibfnamefont {Y.-J.}\ \bibnamefont {Wei}}, \bibinfo {author}
  {\bibfnamefont {D.}~\bibnamefont {Wu}}, \bibinfo {author} {\bibfnamefont
  {M.}~\bibnamefont {Atat{\"u}re}}, \bibinfo {author} {\bibfnamefont
  {C.}~\bibnamefont {Schneider}}, \bibinfo {author} {\bibfnamefont
  {S.}~\bibnamefont {H{\"o}fling}}, \bibinfo {author} {\bibfnamefont
  {M.}~\bibnamefont {Kamp}}, \bibinfo {author} {\bibfnamefont {C.-Y.}\
  \bibnamefont {Lu}}, \ and\ \bibinfo {author} {\bibfnamefont {J.-W.}\
  \bibnamefont {Pan}},\ }\href@noop {} {\bibfield  {journal} {\bibinfo
  {journal} {Nature nanotechnology}\ }\textbf {\bibinfo {volume} {8}},\
  \bibinfo {pages} {213} (\bibinfo {year} {2013})}\BibitemShut {NoStop}%
\bibitem [{\citenamefont {Stevenson}\ \emph {et~al.}(2006)\citenamefont
  {Stevenson}, \citenamefont {Young}, \citenamefont {Atkinson}, \citenamefont
  {Cooper}, \citenamefont {Ritchie},\ and\ \citenamefont
  {Shields}}]{Stevenson2006}%
  \BibitemOpen
  \bibfield  {author} {\bibinfo {author} {\bibfnamefont {R.~M.}\ \bibnamefont
  {Stevenson}}, \bibinfo {author} {\bibfnamefont {R.~J.}\ \bibnamefont
  {Young}}, \bibinfo {author} {\bibfnamefont {P.}~\bibnamefont {Atkinson}},
  \bibinfo {author} {\bibfnamefont {K.}~\bibnamefont {Cooper}}, \bibinfo
  {author} {\bibfnamefont {D.~A.}\ \bibnamefont {Ritchie}}, \ and\ \bibinfo
  {author} {\bibfnamefont {A.~J.}\ \bibnamefont {Shields}},\ }\href@noop {}
  {\bibfield  {journal} {\bibinfo  {journal} {Nature}\ }\textbf {\bibinfo
  {volume} {439}},\ \bibinfo {pages} {179} (\bibinfo {year}
  {2006})}\BibitemShut {NoStop}%
\bibitem [{\citenamefont {Young}\ \emph {et~al.}(2006)\citenamefont {Young},
  \citenamefont {Stevenson}, \citenamefont {Atkinson}, \citenamefont {Cooper},
  \citenamefont {Ritchie},\ and\ \citenamefont {Shields}}]{Young_2006}%
  \BibitemOpen
  \bibfield  {author} {\bibinfo {author} {\bibfnamefont {R.~J.}\ \bibnamefont
  {Young}}, \bibinfo {author} {\bibfnamefont {R.~M.}\ \bibnamefont
  {Stevenson}}, \bibinfo {author} {\bibfnamefont {P.}~\bibnamefont {Atkinson}},
  \bibinfo {author} {\bibfnamefont {K.}~\bibnamefont {Cooper}}, \bibinfo
  {author} {\bibfnamefont {D.~A.}\ \bibnamefont {Ritchie}}, \ and\ \bibinfo
  {author} {\bibfnamefont {A.~J.}\ \bibnamefont {Shields}},\ }\href {\doibase
  10.1088/1367-2630/8/2/029} {\bibfield  {journal} {\bibinfo  {journal} {New
  Journal of Physics}\ }\textbf {\bibinfo {volume} {8}},\ \bibinfo {pages} {29}
  (\bibinfo {year} {2006})}\BibitemShut {NoStop}%
\bibitem [{\citenamefont {Dousse}\ \emph {et~al.}(2010)\citenamefont {Dousse},
  \citenamefont {Suffczy{\'{n}}ski}, \citenamefont {Beveratos}, \citenamefont
  {Krebs}, \citenamefont {Lema{\^i}tre}, \citenamefont {Sagnes}, \citenamefont
  {Bloch}, \citenamefont {Voisin},\ and\ \citenamefont
  {Senellart}}]{Dousse2010}%
  \BibitemOpen
  \bibfield  {author} {\bibinfo {author} {\bibfnamefont {A.}~\bibnamefont
  {Dousse}}, \bibinfo {author} {\bibfnamefont {J.}~\bibnamefont
  {Suffczy{\'{n}}ski}}, \bibinfo {author} {\bibfnamefont {A.}~\bibnamefont
  {Beveratos}}, \bibinfo {author} {\bibfnamefont {O.}~\bibnamefont {Krebs}},
  \bibinfo {author} {\bibfnamefont {A.}~\bibnamefont {Lema{\^i}tre}}, \bibinfo
  {author} {\bibfnamefont {I.}~\bibnamefont {Sagnes}}, \bibinfo {author}
  {\bibfnamefont {J.}~\bibnamefont {Bloch}}, \bibinfo {author} {\bibfnamefont
  {P.}~\bibnamefont {Voisin}}, \ and\ \bibinfo {author} {\bibfnamefont
  {P.}~\bibnamefont {Senellart}},\ }\href {\doibase 10.1038/nature09148}
  {\bibfield  {journal} {\bibinfo  {journal} {Nature}\ }\textbf {\bibinfo
  {volume} {466}},\ \bibinfo {pages} {217} (\bibinfo {year}
  {2010})}\BibitemShut {NoStop}%
\bibitem [{\citenamefont {Basset}\ \emph {et~al.}(2019)\citenamefont {Basset},
  \citenamefont {Rota}, \citenamefont {Schimpf}, \citenamefont {Tedeschi},
  \citenamefont {Zeuner}, \citenamefont {Da~Silva}, \citenamefont {Reindl},
  \citenamefont {Zwiller}, \citenamefont {J{\"o}ns}, \citenamefont {Rastelli}
  \emph {et~al.}}]{PhysRevLett.123.160501}%
  \BibitemOpen
  \bibfield  {author} {\bibinfo {author} {\bibfnamefont {F.~B.}\ \bibnamefont
  {Basset}}, \bibinfo {author} {\bibfnamefont {M.~B.}\ \bibnamefont {Rota}},
  \bibinfo {author} {\bibfnamefont {C.}~\bibnamefont {Schimpf}}, \bibinfo
  {author} {\bibfnamefont {D.}~\bibnamefont {Tedeschi}}, \bibinfo {author}
  {\bibfnamefont {K.~D.}\ \bibnamefont {Zeuner}}, \bibinfo {author}
  {\bibfnamefont {S.~C.}\ \bibnamefont {Da~Silva}}, \bibinfo {author}
  {\bibfnamefont {M.}~\bibnamefont {Reindl}}, \bibinfo {author} {\bibfnamefont
  {V.}~\bibnamefont {Zwiller}}, \bibinfo {author} {\bibfnamefont {K.~D.}\
  \bibnamefont {J{\"o}ns}}, \bibinfo {author} {\bibfnamefont {A.}~\bibnamefont
  {Rastelli}},  \emph {et~al.},\ }\href@noop {} {\bibfield  {journal} {\bibinfo
   {journal} {Physical Review Letters}\ }\textbf {\bibinfo {volume} {123}},\
  \bibinfo {pages} {160501} (\bibinfo {year} {2019})}\BibitemShut {NoStop}%
\bibitem [{\citenamefont {Schwartz}\ \emph {et~al.}(2016)\citenamefont
  {Schwartz}, \citenamefont {Cogan}, \citenamefont {Schmidgall}, \citenamefont
  {Don}, \citenamefont {Gantz}, \citenamefont {Kenneth}, \citenamefont
  {Lindner},\ and\ \citenamefont {Gershoni}}]{doi:10.1126/science.aah4758}%
  \BibitemOpen
  \bibfield  {author} {\bibinfo {author} {\bibfnamefont {I.}~\bibnamefont
  {Schwartz}}, \bibinfo {author} {\bibfnamefont {D.}~\bibnamefont {Cogan}},
  \bibinfo {author} {\bibfnamefont {E.~R.}\ \bibnamefont {Schmidgall}},
  \bibinfo {author} {\bibfnamefont {Y.}~\bibnamefont {Don}}, \bibinfo {author}
  {\bibfnamefont {L.}~\bibnamefont {Gantz}}, \bibinfo {author} {\bibfnamefont
  {O.}~\bibnamefont {Kenneth}}, \bibinfo {author} {\bibfnamefont {N.~H.}\
  \bibnamefont {Lindner}}, \ and\ \bibinfo {author} {\bibfnamefont
  {D.}~\bibnamefont {Gershoni}},\ }\href@noop {} {\bibfield  {journal}
  {\bibinfo  {journal} {Science}\ }\textbf {\bibinfo {volume} {354}},\ \bibinfo
  {pages} {434} (\bibinfo {year} {2016})}\BibitemShut {NoStop}%
\bibitem [{\citenamefont {Lee}\ \emph {et~al.}(2019)\citenamefont {Lee},
  \citenamefont {Villa}, \citenamefont {Bennett}, \citenamefont {Stevenson},
  \citenamefont {Ellis}, \citenamefont {Farrer}, \citenamefont {Ritchie},\ and\
  \citenamefont {Shields}}]{Lee_2019}%
  \BibitemOpen
  \bibfield  {author} {\bibinfo {author} {\bibfnamefont {J.~P.}\ \bibnamefont
  {Lee}}, \bibinfo {author} {\bibfnamefont {B.}~\bibnamefont {Villa}}, \bibinfo
  {author} {\bibfnamefont {A.~J.}\ \bibnamefont {Bennett}}, \bibinfo {author}
  {\bibfnamefont {R.~M.}\ \bibnamefont {Stevenson}}, \bibinfo {author}
  {\bibfnamefont {D.~J.~P.}\ \bibnamefont {Ellis}}, \bibinfo {author}
  {\bibfnamefont {I.}~\bibnamefont {Farrer}}, \bibinfo {author} {\bibfnamefont
  {D.~A.}\ \bibnamefont {Ritchie}}, \ and\ \bibinfo {author} {\bibfnamefont
  {A.~J.}\ \bibnamefont {Shields}},\ }\href {\doibase 10.1088/2058-9565/ab0a9b}
  {\bibfield  {journal} {\bibinfo  {journal} {Quantum Science and Technology}\
  }\textbf {\bibinfo {volume} {4}},\ \bibinfo {pages} {025011} (\bibinfo {year}
  {2019})}\BibitemShut {NoStop}%
\bibitem [{\citenamefont {Appel}\ \emph {et~al.}(2022)\citenamefont {Appel},
  \citenamefont {Tiranov}, \citenamefont {Pabst}, \citenamefont {Chan},
  \citenamefont {Starup}, \citenamefont {Wang}, \citenamefont {Midolo},
  \citenamefont {Tiurev}, \citenamefont {Scholz}, \citenamefont {Wieck} \emph
  {et~al.}}]{PhysRevLett.128.233602}%
  \BibitemOpen
  \bibfield  {author} {\bibinfo {author} {\bibfnamefont {M.~H.}\ \bibnamefont
  {Appel}}, \bibinfo {author} {\bibfnamefont {A.}~\bibnamefont {Tiranov}},
  \bibinfo {author} {\bibfnamefont {S.}~\bibnamefont {Pabst}}, \bibinfo
  {author} {\bibfnamefont {M.~L.}\ \bibnamefont {Chan}}, \bibinfo {author}
  {\bibfnamefont {C.}~\bibnamefont {Starup}}, \bibinfo {author} {\bibfnamefont
  {Y.}~\bibnamefont {Wang}}, \bibinfo {author} {\bibfnamefont {L.}~\bibnamefont
  {Midolo}}, \bibinfo {author} {\bibfnamefont {K.}~\bibnamefont {Tiurev}},
  \bibinfo {author} {\bibfnamefont {S.}~\bibnamefont {Scholz}}, \bibinfo
  {author} {\bibfnamefont {A.~D.}\ \bibnamefont {Wieck}},  \emph {et~al.},\
  }\href@noop {} {\bibfield  {journal} {\bibinfo  {journal} {Physical Review
  Letters}\ }\textbf {\bibinfo {volume} {128}},\ \bibinfo {pages} {233602}
  (\bibinfo {year} {2022})}\BibitemShut {NoStop}%
\bibitem [{\citenamefont {Ding}\ \emph {et~al.}(2016)\citenamefont {Ding},
  \citenamefont {He}, \citenamefont {Duan}, \citenamefont {Gregersen},
  \citenamefont {Chen}, \citenamefont {Unsleber}, \citenamefont {Maier},
  \citenamefont {Schneider}, \citenamefont {Kamp}, \citenamefont {H{\"o}fling},
  \citenamefont {Lu},\ and\ \citenamefont {Pan}}]{Ding:2016fk}%
  \BibitemOpen
  \bibfield  {author} {\bibinfo {author} {\bibfnamefont {X.}~\bibnamefont
  {Ding}}, \bibinfo {author} {\bibfnamefont {Y.}~\bibnamefont {He}}, \bibinfo
  {author} {\bibfnamefont {Z.~C.}\ \bibnamefont {Duan}}, \bibinfo {author}
  {\bibfnamefont {N.}~\bibnamefont {Gregersen}}, \bibinfo {author}
  {\bibfnamefont {M.~C.}\ \bibnamefont {Chen}}, \bibinfo {author}
  {\bibfnamefont {S.}~\bibnamefont {Unsleber}}, \bibinfo {author}
  {\bibfnamefont {S.}~\bibnamefont {Maier}}, \bibinfo {author} {\bibfnamefont
  {C.}~\bibnamefont {Schneider}}, \bibinfo {author} {\bibfnamefont
  {M.}~\bibnamefont {Kamp}}, \bibinfo {author} {\bibfnamefont {S.}~\bibnamefont
  {H{\"o}fling}}, \bibinfo {author} {\bibfnamefont {C.-Y.}\ \bibnamefont {Lu}},
  \ and\ \bibinfo {author} {\bibfnamefont {J.-W.}\ \bibnamefont {Pan}},\ }\href
  {http://link.aps.org/doi/10.1103/PhysRevLett.116.020401} {\bibfield
  {journal} {\bibinfo  {journal} {Physical Review Letters}\ }\textbf {\bibinfo
  {volume} {116}},\ \bibinfo {pages} {020401} (\bibinfo {year}
  {2016})}\BibitemShut {NoStop}%
\bibitem [{\citenamefont {Wang}\ \emph {et~al.}(2019)\citenamefont {Wang},
  \citenamefont {He}, \citenamefont {Chung}, \citenamefont {Hu}, \citenamefont
  {Yu}, \citenamefont {Chen}, \citenamefont {Ding}, \citenamefont {Chen},
  \citenamefont {Qin}, \citenamefont {Yang}, \citenamefont {Liu}, \citenamefont
  {Duan}, \citenamefont {Li}, \citenamefont {Gerhardt}, \citenamefont
  {Winkler}, \citenamefont {Jurkat}, \citenamefont {Wang}, \citenamefont
  {Gregersen}, \citenamefont {Huo}, \citenamefont {Dai}, \citenamefont {Yu},
  \citenamefont {H{\"o}fling}, \citenamefont {Lu},\ and\ \citenamefont
  {Pan}}]{Wang2019}%
  \BibitemOpen
  \bibfield  {author} {\bibinfo {author} {\bibfnamefont {H.}~\bibnamefont
  {Wang}}, \bibinfo {author} {\bibfnamefont {Y.-M.}\ \bibnamefont {He}},
  \bibinfo {author} {\bibfnamefont {T.-H.}\ \bibnamefont {Chung}}, \bibinfo
  {author} {\bibfnamefont {H.}~\bibnamefont {Hu}}, \bibinfo {author}
  {\bibfnamefont {Y.}~\bibnamefont {Yu}}, \bibinfo {author} {\bibfnamefont
  {S.}~\bibnamefont {Chen}}, \bibinfo {author} {\bibfnamefont {X.}~\bibnamefont
  {Ding}}, \bibinfo {author} {\bibfnamefont {M.-C.}\ \bibnamefont {Chen}},
  \bibinfo {author} {\bibfnamefont {J.}~\bibnamefont {Qin}}, \bibinfo {author}
  {\bibfnamefont {X.}~\bibnamefont {Yang}}, \bibinfo {author} {\bibfnamefont
  {R.-Z.}\ \bibnamefont {Liu}}, \bibinfo {author} {\bibfnamefont {Z.-C.}\
  \bibnamefont {Duan}}, \bibinfo {author} {\bibfnamefont {J.-P.}\ \bibnamefont
  {Li}}, \bibinfo {author} {\bibfnamefont {S.}~\bibnamefont {Gerhardt}},
  \bibinfo {author} {\bibfnamefont {K.}~\bibnamefont {Winkler}}, \bibinfo
  {author} {\bibfnamefont {J.}~\bibnamefont {Jurkat}}, \bibinfo {author}
  {\bibfnamefont {L.-J.}\ \bibnamefont {Wang}}, \bibinfo {author}
  {\bibfnamefont {N.}~\bibnamefont {Gregersen}}, \bibinfo {author}
  {\bibfnamefont {Y.-H.}\ \bibnamefont {Huo}}, \bibinfo {author} {\bibfnamefont
  {Q.}~\bibnamefont {Dai}}, \bibinfo {author} {\bibfnamefont {S.}~\bibnamefont
  {Yu}}, \bibinfo {author} {\bibfnamefont {S.}~\bibnamefont {H{\"o}fling}},
  \bibinfo {author} {\bibfnamefont {C.-Y.}\ \bibnamefont {Lu}}, \ and\ \bibinfo
  {author} {\bibfnamefont {J.-W.}\ \bibnamefont {Pan}},\ }\href {\doibase
  10.1038/s41566-019-0494-3} {\bibfield  {journal} {\bibinfo  {journal} {Nature
  Photonics}\ }\textbf {\bibinfo {volume} {13}},\ \bibinfo {pages} {770}
  (\bibinfo {year} {2019})}\BibitemShut {NoStop}%
\bibitem [{\citenamefont {Santori}\ \emph {et~al.}(2002)\citenamefont
  {Santori}, \citenamefont {Fattal}, \citenamefont {Vu{\v{c}}kovi{\'{c}}},
  \citenamefont {Solomon},\ and\ \citenamefont {Yamamoto}}]{Santori2002}%
  \BibitemOpen
  \bibfield  {author} {\bibinfo {author} {\bibfnamefont {C.}~\bibnamefont
  {Santori}}, \bibinfo {author} {\bibfnamefont {D.}~\bibnamefont {Fattal}},
  \bibinfo {author} {\bibfnamefont {J.}~\bibnamefont {Vu{\v{c}}kovi{\'{c}}}},
  \bibinfo {author} {\bibfnamefont {G.~S.}\ \bibnamefont {Solomon}}, \ and\
  \bibinfo {author} {\bibfnamefont {Y.}~\bibnamefont {Yamamoto}},\ }\href
  {\doibase 10.1038/nature01086} {\bibfield  {journal} {\bibinfo  {journal}
  {Nature}\ }\textbf {\bibinfo {volume} {419}},\ \bibinfo {pages} {594}
  (\bibinfo {year} {2002})}\BibitemShut {NoStop}%
\bibitem [{\citenamefont {Bennett}\ \emph {et~al.}(2016)\citenamefont
  {Bennett}, \citenamefont {Lee}, \citenamefont {Ellis}, \citenamefont
  {Farrer}, \citenamefont {Ritchie},\ and\ \citenamefont
  {Shields}}]{Bennett2016}%
  \BibitemOpen
  \bibfield  {author} {\bibinfo {author} {\bibfnamefont {A.~J.}\ \bibnamefont
  {Bennett}}, \bibinfo {author} {\bibfnamefont {J.~P.}\ \bibnamefont {Lee}},
  \bibinfo {author} {\bibfnamefont {D.~J.~P.}\ \bibnamefont {Ellis}}, \bibinfo
  {author} {\bibfnamefont {I.}~\bibnamefont {Farrer}}, \bibinfo {author}
  {\bibfnamefont {D.~A.}\ \bibnamefont {Ritchie}}, \ and\ \bibinfo {author}
  {\bibfnamefont {A.~J.}\ \bibnamefont {Shields}},\ }\href {\doibase
  10.1038/nnano.2016.113} {\bibfield  {journal} {\bibinfo  {journal} {Nature
  Nanotechnology}\ }\textbf {\bibinfo {volume} {11}},\ \bibinfo {pages} {857}
  (\bibinfo {year} {2016})}\BibitemShut {NoStop}%
\bibitem [{\citenamefont {Schneider}\ \emph {et~al.}(2016)\citenamefont
  {Schneider}, \citenamefont {Gold}, \citenamefont {Reitzenstein},
  \citenamefont {H{\"o}fling},\ and\ \citenamefont {Kamp}}]{Schneider2016}%
  \BibitemOpen
  \bibfield  {author} {\bibinfo {author} {\bibfnamefont {C.}~\bibnamefont
  {Schneider}}, \bibinfo {author} {\bibfnamefont {P.}~\bibnamefont {Gold}},
  \bibinfo {author} {\bibfnamefont {S.}~\bibnamefont {Reitzenstein}}, \bibinfo
  {author} {\bibfnamefont {S.}~\bibnamefont {H{\"o}fling}}, \ and\ \bibinfo
  {author} {\bibfnamefont {M.}~\bibnamefont {Kamp}},\ }\href {\doibase
  10.1007/s00340-015-6283-x} {\bibfield  {journal} {\bibinfo  {journal}
  {Applied Physics B}\ }\textbf {\bibinfo {volume} {122}},\ \bibinfo {pages}
  {19} (\bibinfo {year} {2016})}\BibitemShut {NoStop}%
\bibitem [{\citenamefont {Cho}\ \emph {et~al.}(2010)\citenamefont {Cho},
  \citenamefont {Kim}, \citenamefont {Song}, \citenamefont {Choi},\ and\
  \citenamefont {Lee}}]{CHO20101955}%
  \BibitemOpen
  \bibfield  {author} {\bibinfo {author} {\bibfnamefont {N.}~\bibnamefont
  {Cho}}, \bibinfo {author} {\bibfnamefont {K.}~\bibnamefont {Kim}}, \bibinfo
  {author} {\bibfnamefont {J.}~\bibnamefont {Song}}, \bibinfo {author}
  {\bibfnamefont {W.}~\bibnamefont {Choi}}, \ and\ \bibinfo {author}
  {\bibfnamefont {J.}~\bibnamefont {Lee}},\ }\href {\doibase
  https://doi.org/10.1016/j.ssc.2010.05.010} {\bibfield  {journal} {\bibinfo
  {journal} {Solid State Communications}\ }\textbf {\bibinfo {volume} {150}},\
  \bibinfo {pages} {1955} (\bibinfo {year} {2010})}\BibitemShut {NoStop}%
\bibitem [{\citenamefont {Androvitsaneas}\ \emph {et~al.}(2016)\citenamefont
  {Androvitsaneas}, \citenamefont {Young}, \citenamefont {Schneider},
  \citenamefont {Maier}, \citenamefont {Kamp}, \citenamefont {H{\"o}fling},
  \citenamefont {Knauer}, \citenamefont {Harbord}, \citenamefont {Hu},
  \citenamefont {Rarity},\ and\ \citenamefont
  {Oulton}}]{Androvitsaneas:2016kx}%
  \BibitemOpen
  \bibfield  {author} {\bibinfo {author} {\bibfnamefont {P.}~\bibnamefont
  {Androvitsaneas}}, \bibinfo {author} {\bibfnamefont {A.~B.}\ \bibnamefont
  {Young}}, \bibinfo {author} {\bibfnamefont {C.}~\bibnamefont {Schneider}},
  \bibinfo {author} {\bibfnamefont {S.}~\bibnamefont {Maier}}, \bibinfo
  {author} {\bibfnamefont {M.}~\bibnamefont {Kamp}}, \bibinfo {author}
  {\bibfnamefont {S.}~\bibnamefont {H{\"o}fling}}, \bibinfo {author}
  {\bibfnamefont {S.}~\bibnamefont {Knauer}}, \bibinfo {author} {\bibfnamefont
  {E.}~\bibnamefont {Harbord}}, \bibinfo {author} {\bibfnamefont {C.~Y.}\
  \bibnamefont {Hu}}, \bibinfo {author} {\bibfnamefont {J.~G.}\ \bibnamefont
  {Rarity}}, \ and\ \bibinfo {author} {\bibfnamefont {R.}~\bibnamefont
  {Oulton}},\ }\href {http://link.aps.org/doi/10.1103/PhysRevB.93.241409}
  {\bibfield  {journal} {\bibinfo  {journal} {Physical Review B}\ }\textbf
  {\bibinfo {volume} {93}},\ \bibinfo {pages} {241409} (\bibinfo {year}
  {2016})}\BibitemShut {NoStop}%
\bibitem [{\citenamefont {Gin{\'e}s}\ \emph {et~al.}(2022)\citenamefont
  {Gin{\'e}s}, \citenamefont {Mocza{\l}a-Dusanowska}, \citenamefont {Dlaka},
  \citenamefont {Ho{\v{s}}{\'a}k}, \citenamefont {Gonzales-Ureta},
  \citenamefont {Lee}, \citenamefont {Je{\v{z}}ek}, \citenamefont {Harbord},
  \citenamefont {Oulton}, \citenamefont {H{\"o}fling} \emph
  {et~al.}}]{PhysRevLett.129.033601}%
  \BibitemOpen
  \bibfield  {author} {\bibinfo {author} {\bibfnamefont {L.}~\bibnamefont
  {Gin{\'e}s}}, \bibinfo {author} {\bibfnamefont {M.}~\bibnamefont
  {Mocza{\l}a-Dusanowska}}, \bibinfo {author} {\bibfnamefont {D.}~\bibnamefont
  {Dlaka}}, \bibinfo {author} {\bibfnamefont {R.}~\bibnamefont
  {Ho{\v{s}}{\'a}k}}, \bibinfo {author} {\bibfnamefont {J.~R.}\ \bibnamefont
  {Gonzales-Ureta}}, \bibinfo {author} {\bibfnamefont {J.}~\bibnamefont {Lee}},
  \bibinfo {author} {\bibfnamefont {M.}~\bibnamefont {Je{\v{z}}ek}}, \bibinfo
  {author} {\bibfnamefont {E.}~\bibnamefont {Harbord}}, \bibinfo {author}
  {\bibfnamefont {R.}~\bibnamefont {Oulton}}, \bibinfo {author} {\bibfnamefont
  {S.}~\bibnamefont {H{\"o}fling}},  \emph {et~al.},\ }\href@noop {} {\bibfield
   {journal} {\bibinfo  {journal} {Physical Review Letters}\ }\textbf {\bibinfo
  {volume} {129}},\ \bibinfo {pages} {033601} (\bibinfo {year}
  {2022})}\BibitemShut {NoStop}%
\bibitem [{\citenamefont {Kuhlmann}\ \emph {et~al.}(2013)\citenamefont
  {Kuhlmann}, \citenamefont {Houel}, \citenamefont {Brunner}, \citenamefont
  {Ludwig}, \citenamefont {Reuter}, \citenamefont {Wieck},\ and\ \citenamefont
  {Warburton}}]{kuhlmandark}%
  \BibitemOpen
  \bibfield  {author} {\bibinfo {author} {\bibfnamefont {A.~V.}\ \bibnamefont
  {Kuhlmann}}, \bibinfo {author} {\bibfnamefont {J.}~\bibnamefont {Houel}},
  \bibinfo {author} {\bibfnamefont {D.}~\bibnamefont {Brunner}}, \bibinfo
  {author} {\bibfnamefont {A.}~\bibnamefont {Ludwig}}, \bibinfo {author}
  {\bibfnamefont {D.}~\bibnamefont {Reuter}}, \bibinfo {author} {\bibfnamefont
  {A.~D.}\ \bibnamefont {Wieck}}, \ and\ \bibinfo {author} {\bibfnamefont
  {R.~J.}\ \bibnamefont {Warburton}},\ }\href {\doibase
  http://dx.doi.org/10.1063/1.4813879} {\bibfield  {journal} {\bibinfo
  {journal} {Review of Scientific Instruments}\ }\textbf {\bibinfo {volume}
  {84}},\ \bibinfo {eid} {073905} (\bibinfo {year} {2013})}\BibitemShut
  {NoStop}%
\bibitem [{\citenamefont {Brown}\ and\ \citenamefont {Twiss}(1954)}]{hbt}%
  \BibitemOpen
  \bibfield  {author} {\bibinfo {author} {\bibfnamefont {R.~H.}\ \bibnamefont
  {Brown}}\ and\ \bibinfo {author} {\bibfnamefont {R.~Q.}\ \bibnamefont
  {Twiss}},\ }\href@noop {} {\bibfield  {journal} {\bibinfo  {journal} {The
  London, Edinburgh, and Dublin Philosophical Magazine and Journal of Science}\
  }\textbf {\bibinfo {volume} {45}},\ \bibinfo {pages} {663} (\bibinfo {year}
  {1954})}\BibitemShut {NoStop}%
\bibitem [{\citenamefont {Ollivier}\ \emph {et~al.}(2021)\citenamefont
  {Ollivier}, \citenamefont {Thomas}, \citenamefont {Wein}, \citenamefont
  {de~Buy~Wenniger}, \citenamefont {Coste}, \citenamefont {Loredo},
  \citenamefont {Somaschi}, \citenamefont {Harouri}, \citenamefont {Lemaitre},
  \citenamefont {Sagnes} \emph {et~al.}}]{hom_cor}%
  \BibitemOpen
  \bibfield  {author} {\bibinfo {author} {\bibfnamefont {H.}~\bibnamefont
  {Ollivier}}, \bibinfo {author} {\bibfnamefont {S.}~\bibnamefont {Thomas}},
  \bibinfo {author} {\bibfnamefont {S.}~\bibnamefont {Wein}}, \bibinfo {author}
  {\bibfnamefont {I.~M.}\ \bibnamefont {de~Buy~Wenniger}}, \bibinfo {author}
  {\bibfnamefont {N.}~\bibnamefont {Coste}}, \bibinfo {author} {\bibfnamefont
  {J.}~\bibnamefont {Loredo}}, \bibinfo {author} {\bibfnamefont
  {N.}~\bibnamefont {Somaschi}}, \bibinfo {author} {\bibfnamefont
  {A.}~\bibnamefont {Harouri}}, \bibinfo {author} {\bibfnamefont
  {A.}~\bibnamefont {Lemaitre}}, \bibinfo {author} {\bibfnamefont
  {I.}~\bibnamefont {Sagnes}},  \emph {et~al.},\ }\href@noop {} {\bibfield
  {journal} {\bibinfo  {journal} {Physical Review Letters}\ }\textbf {\bibinfo
  {volume} {126}},\ \bibinfo {pages} {063602} (\bibinfo {year}
  {2021})}\BibitemShut {NoStop}%
\bibitem [{\citenamefont {Gro{\ss}e}\ \emph {et~al.}(2020)\citenamefont
  {Gro{\ss}e}, \citenamefont {von Helversen}, \citenamefont {Koulas-Simos},
  \citenamefont {Hermann},\ and\ \citenamefont
  {Reitzenstein}}]{doi:10.1063/5.0013718}%
  \BibitemOpen
  \bibfield  {author} {\bibinfo {author} {\bibfnamefont {J.}~\bibnamefont
  {Gro{\ss}e}}, \bibinfo {author} {\bibfnamefont {M.}~\bibnamefont {von
  Helversen}}, \bibinfo {author} {\bibfnamefont {A.}~\bibnamefont
  {Koulas-Simos}}, \bibinfo {author} {\bibfnamefont {M.}~\bibnamefont
  {Hermann}}, \ and\ \bibinfo {author} {\bibfnamefont {S.}~\bibnamefont
  {Reitzenstein}},\ }\href@noop {} {\bibfield  {journal} {\bibinfo  {journal}
  {APL Photonics}\ }\textbf {\bibinfo {volume} {5}},\ \bibinfo {pages} {096107}
  (\bibinfo {year} {2020})}\BibitemShut {NoStop}%
\end{thebibliography}%
\end{document}